\documentclass{aa}
\usepackage{psfig}

\begin{document}
\thesaurus{01
          (11.03.01;
           11.17.4: Cloverleaf;
           12.03.3
           12.07.01
           12.12.1)}
\title{
Unveiling the nature of the Cloverleaf lens-system:
HST/NICMOS-2 observations
\thanks{
Based on observations obtained with
the
NASA/ESA Hubble Space Telescope, obtained from the data archive at the
Space Telescope Institute. STScI is operated by the Association of
Universities for Research in Astronomy, Inc. under the NASA contract 
NAS 5-26555.
}
}
\author{J.-P. Kneib\inst{1}
        D. Alloin\inst{2}
	R. Pell\'o\inst{1}
}
\offprints{D. Alloin
           alloin@discovery.saclay.cea.fr}
\institute{
Observatoire Midi-Pyr\'en\'ees, CNRS-UMR5572, 14
Av. Edouard Belin, 31400 Toulouse, France
\and 
CNRS-URA2052, Service d'Astrophysique, CE Saclay, l'Orme des Merisiers,
91191 
Gif-sur -Yvette Cedex, France
}
\date{Received April XX 1998, Accepted Yyy ZZ }

\authorrunning{J.-P. Kneib et al, 1998}
\titlerunning{NICMOS-2 observation of the Cloverleaf}
\maketitle
\markboth{J.-P. Kneib et al, 1998}{NICMOS-2 observation of the Cloverleaf}

\begin{abstract}
We present new elements in the identification of the lens-system
producing the 4 images of the BAL quasar H1413+117, based on the 
recent HST/NICMOS-2/F160W observations.
After a careful PSF subtraction of the  
4 images of the quasar, the residual H$_{F160W}$ image
reveals the presence of a faint object ($H\sim 20.5$)
within the region enclosed by
the 4 quasar images. This object corresponds to a single
galaxy: the primary lens of the lens-system. We also identify  
the galaxies around the Cloverleaf which had been proposed
to belong to a galaxy cluster/group at high
redshift (Kneib et al 1998): 
the other component in the lens-system that provides
the additional ``external" shear.
For these galaxies, we have derived a likely
redshift based upon their R$_{F702W}$, I$_{F814W}$ 
and H$_{F160W}$ magnitudes.
We find that most of them are consistent with 
belonging to a galaxy cluster/group
with mean redshift $\overline{z}=0.9 \pm 0.1$. 
Furthermore we detect 2 very red objects ($I-H\sim 4$): 
the faintest one has no observed optical (R$_{F702W}$ and 
I$_{F814W}$) counterpart, while the brightest has a predicted 
redshift around $z\sim 2$, and may be identified with one of the
Cloverleaf absorbers.
This gravitational-lens system constitutes an excellent target for IR
imaging/spectroscopy with the new generation of 8m ground-based telescopes.

\end{abstract}

\section{Introduction}

The excellent quality and broad wavelength coverage
of current observations of gravitational lensing systems
have allowed to unveil part of their mysteries. 
The primary lens is often detected. The immediate
surrounding of the multiply imaged quasars/galaxies is studied
in great detail and generally shows some galaxy clustering ({\it e.g.}
Tonry 1998, Hjorth \& Kneib 1998) or even sometimes X-ray 
cluster emissions (Hattori et al 1997, Chartas et al 1998).
Furthermore, the measure of a time-delay in Q0957+561/PG1115 (Kundic et al
1997, Schechter et al 1997) has strengthened the interest of a detailed 
study of these gravitational lens systems in order to use them as a 
cosmological tool.

Since the identification by Magain et al (1988) of the quadrupole
lens-system called the Cloverleaf, 4 images of the BAL quasar
H1413+117 at z=2.558, many efforts have been dedicated to a direct
search of the lens or of elements of the lens-system. Early
models of the lens-system have involved one
or two galaxy-lenses very close to the line-of-sight toward the quasar
(Kayser et al, 1990). A more recent analysis showed that
an external shear was needed to model this system correctly 
(Keeton, Kochaneck \& Seljak 1997), and indeed it is probably 
related to the existence of an overdensity of galaxies nearby,
as detected by Kneib et al (1998). 
 
The lensing geometry, 
amplification and time-delays are quite sensitive to the parameters of the
galaxy-lens expected to be located amid the 4 images of the quasar. A
positive detection of the galaxy-lens would bring stringent
constraints in the modeling of the lens-system.  However, despite
relatively deep searches, either in R and I imaging with
the HST (Turnshek et al, 1997; Kneib et al, 1998) or K imaging with 
the Keck telescope 
(Lawrence et al 1996), the lensing galaxy has not been detected.

The  galaxy cluster recently 
revealed near the Cloverleaf (Kneib et al, 1998)
was assumed to be at $z\sim 1.7$ as this corresponds to the mean value of the
narrow absorption line-systems observed in the quasar spectra (z=1.44, 1.66,
1.87, 2.07 and 2.09: Turnshek et al, 1988; 
Magain et al, 1988, Monier et al 1998). 
Combining the IR image with the R$_{F702W}$ and I$_{F814W}$
WFPC-2 images of this system can allow to
estimate the likely redshift for the faint galaxies surrounding the Cloverleaf.
The recently acquired HST/NICMOS-2/F160W observation consists of a unique
dataset to help answer both the existence of the lensing galaxy and
to constrain the distance of the nearby galaxies.

The HST/NICMOS-2 data are presented in Sect.2, while in Sect. 3 we discuss the 
identification of the lensing galaxy after the PSF subtraction of the
4 quasar images. The redshift estimates (derived from photometry)
of the surrounding
galaxies are explained in Sect.4. The discussion and concluding remarks are 
provided in Sect.5.
Throughout this paper we use $H_0=$50 km/s/Mpc and $\Omega_0=$1.

\section{The HST/NICMOS-2 data}

\begin{figure}
\psfig{file=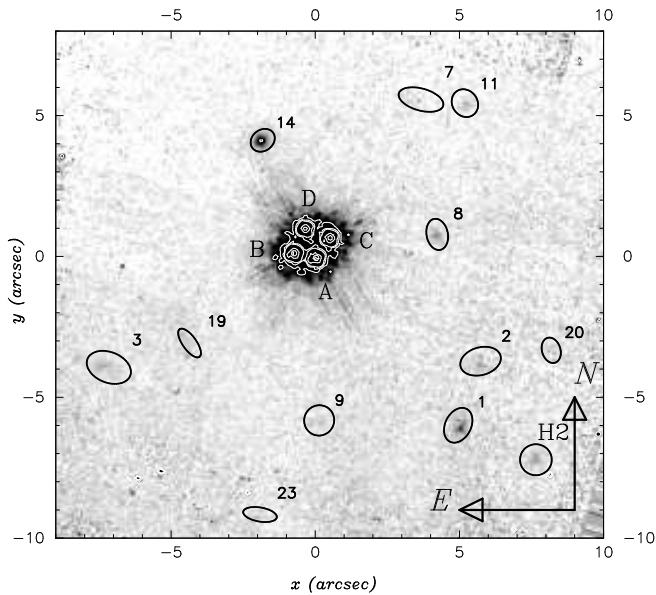,width=0.52\textwidth}
\caption{
Final NICMOS-2/F160W image of the Cloverleaf. Ellipses indicate  
the objects detected in the WFPC2/F814W image (following the numbering
of Kneib et al 1998).
\label{fig:image}
}
\end{figure}

We retrieved the NICMOS-2/F160W observations of the Cloverleaf which
do not hold proprietary rights from the HST/STSci archive:
http://archive.stsci.edu/cgi-bin/hst.
These data are part of a much larger
dataset (PI: E. Falco) with the goal of studying
the statistical properties of a sample of gravitational lens systems.
The Cloverleaf data were taken on December 28, 1997 with a
total exposure time of 4$\times$640 seconds.
At the date of the observations, the NICMOS-2 pixel 
size {\it (x,y)} was 0.07603" x 0.07534"
({\it c.f.} the NICMOS instrument WEB-page).
The NICMOS data come in 3 different formats: raw (files *\_raw.fits), 
calibrated (files *\_cal.fits) and mosaiced (files *\_mos.fits).
The raw data set is made of 4 sets of 19 rasters. 
The calibrated data correspond to 4 flux calibrated
images, each being a combination of the 19 flat-field corrected images. 
We made our own mosaiced image by combining the  4 flux calibrated
images shifted with a non-integer number of pixels (Figure \ref{fig:image}).
The image quality measured on the final mosaiced image is 0.145" (FWHM).

The photometric calibration in the Vega system
was determined from the PHOTNU image keyword following the NICMOS handbook
recipe (PHOTNU=2.3857 10$^{-6}$). 
The software SExtractor 1.2 (Bertin \& Arnouts 1996) was used to
detect and to derive the photometry of faint sources,
the detection parameters used are: 15 contiguous pixels over 1.5$\sigma$ of the
sky-background after a convolution with a 3$\times$3 top-hat filter.
Photometry and position relative to the brightest quasar image (A)
of the different objects is summarized in Table~\ref{tab:gal}.

\section{The 4 quasar images and the primary galaxy-lens}


In order to reveal the primary galaxy-lens amid the 4 quasar images, we 
have subtracted modeled
PSFs of the NICMOS-2 camera. We used the TinyTim V4.4 software
(Krist \& Hook 1997)
to model the NICMOS-2 PSF through the F160W filter
(assuming a flat spectrum for the quasar through this waveband) at 
each of the 4 positions of the quasar images and in each of the  
4 calibrated frames. We oversampled the PSF with 8$\times$8 subpixels
per NICMOS-2 pixel (in order to sample correctly the PSFs and position
them accurately). 
The positions of the 4 images of the Cloverleaf given in Turnshek et al (1997)
were used for the PSF subtraction (we also independantly fitted the
position directly from the NICMOS image but it did not improved the
substraction), while the flux ratios were adjusted in order to minimize
the {\it mean} residuals at the 4 quasar image 
locations. This procedure was applied 
to each of the 4 calibrated images, allowing us to derive the 
magnitudes and flux
ratios of the 4 quasar images in H$_{F160W}$ (Table \ref{tab:qso}),
and it also gives independant errors on the photometry of the different images.
We found systematic residual patterns for each of the 4
PSF-substracted quasar images -- as it can be seen in Figure 2 --
but these patterns do not corresponds to the location of the detected
lensing galaxy. Hence we could not check for the existence of 
an Einstein ring.  Flux ratios in H
differ significantly from the ones in U$_{F336W}$, R$_{F702W}$ and 
I$_{F814W}$, particularly for 
image D. As proposed by Turnshek et al (1997), dust extinction can
easily explain the different flux ratios. However microlensing on the
D image is also very likely, as it is the closest image to the
detected lensing-galaxy.

\begin{figure}
\begin{minipage}{0.24\textwidth}
\psfig{file=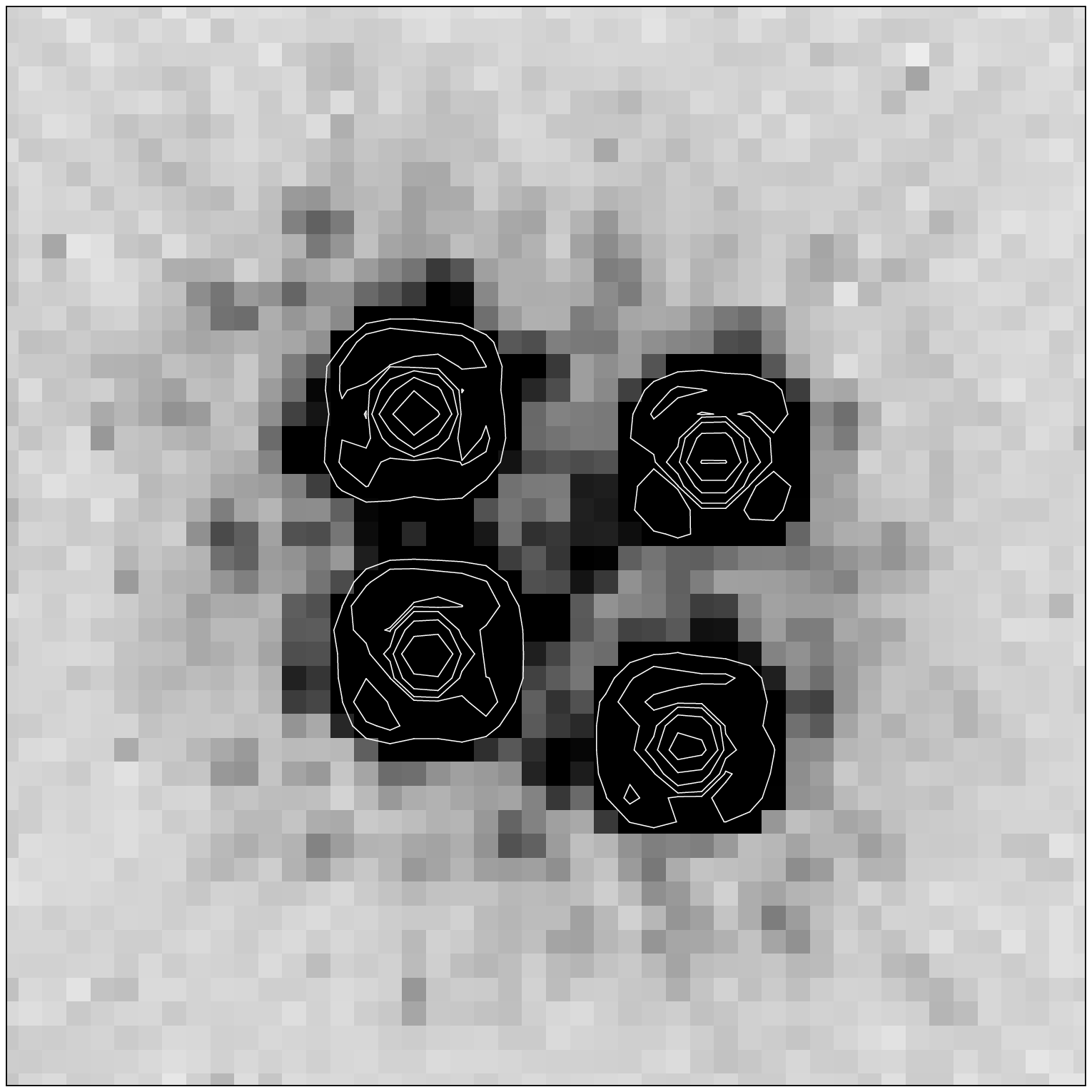,width=\textwidth}
\end{minipage}
\begin{minipage}{0.24\textwidth}
\psfig{file=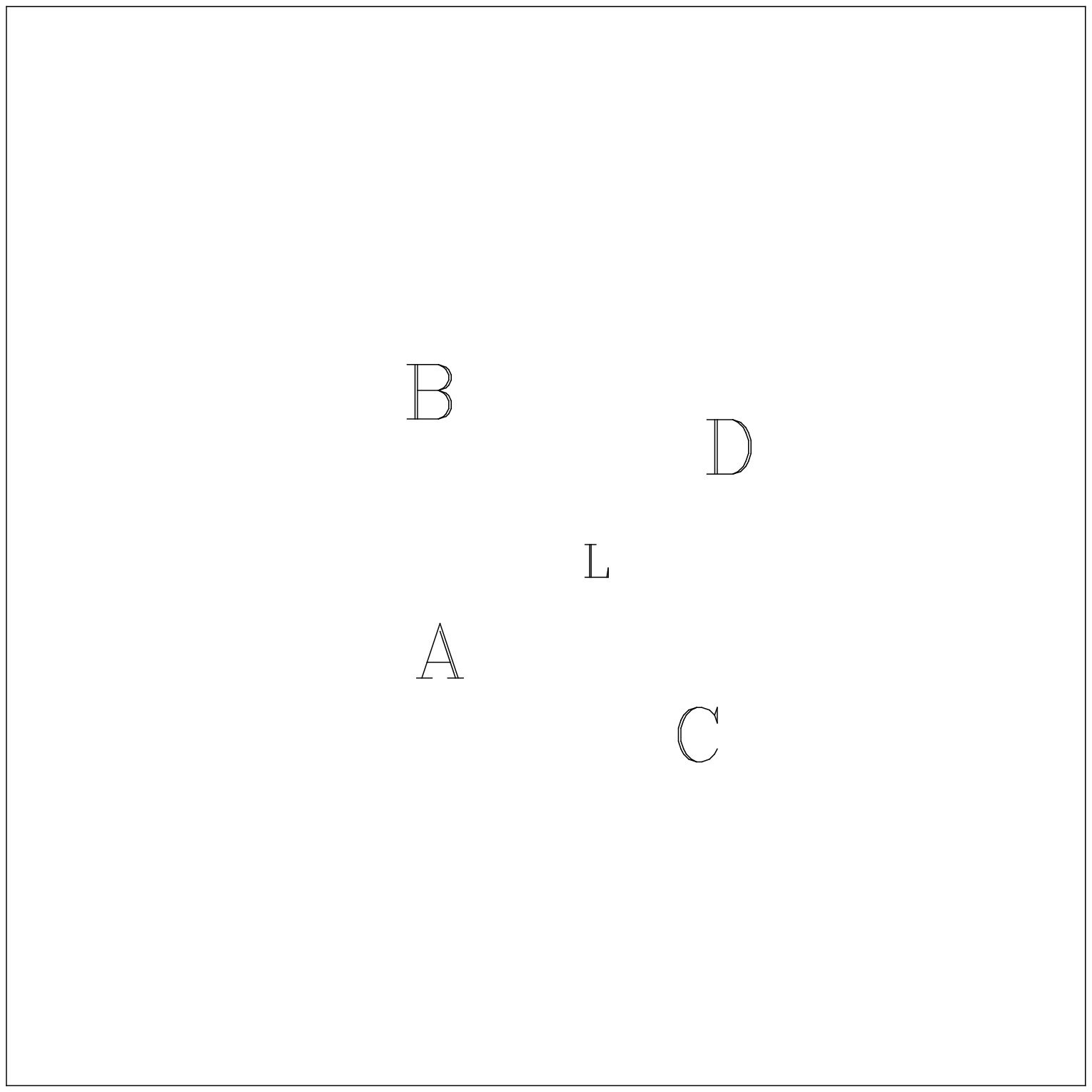,width=\textwidth}
\end{minipage}

\begin{minipage}{0.24\textwidth}
\psfig{file=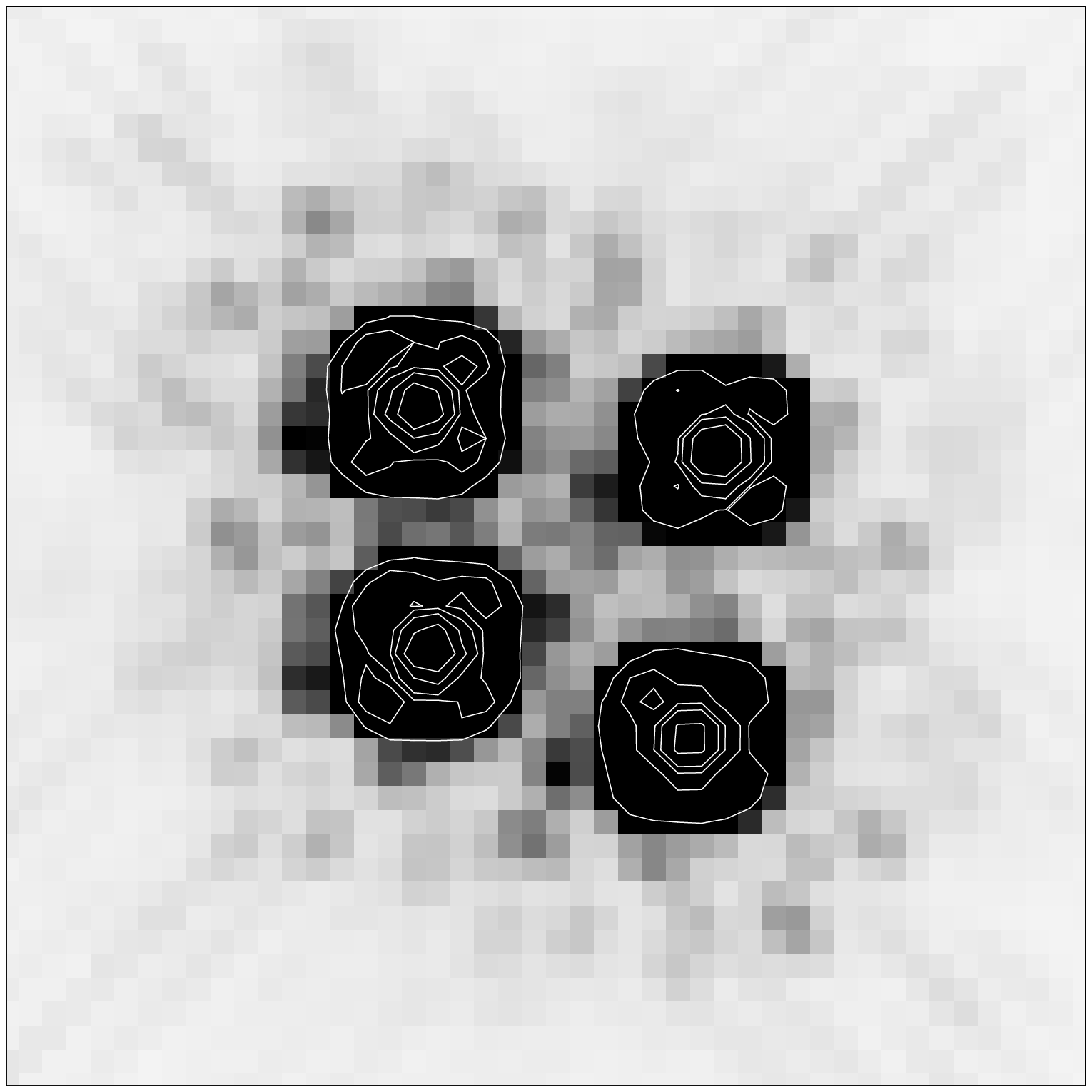,width=\textwidth}
\end{minipage}
\begin{minipage}{0.24\textwidth}
\psfig{file=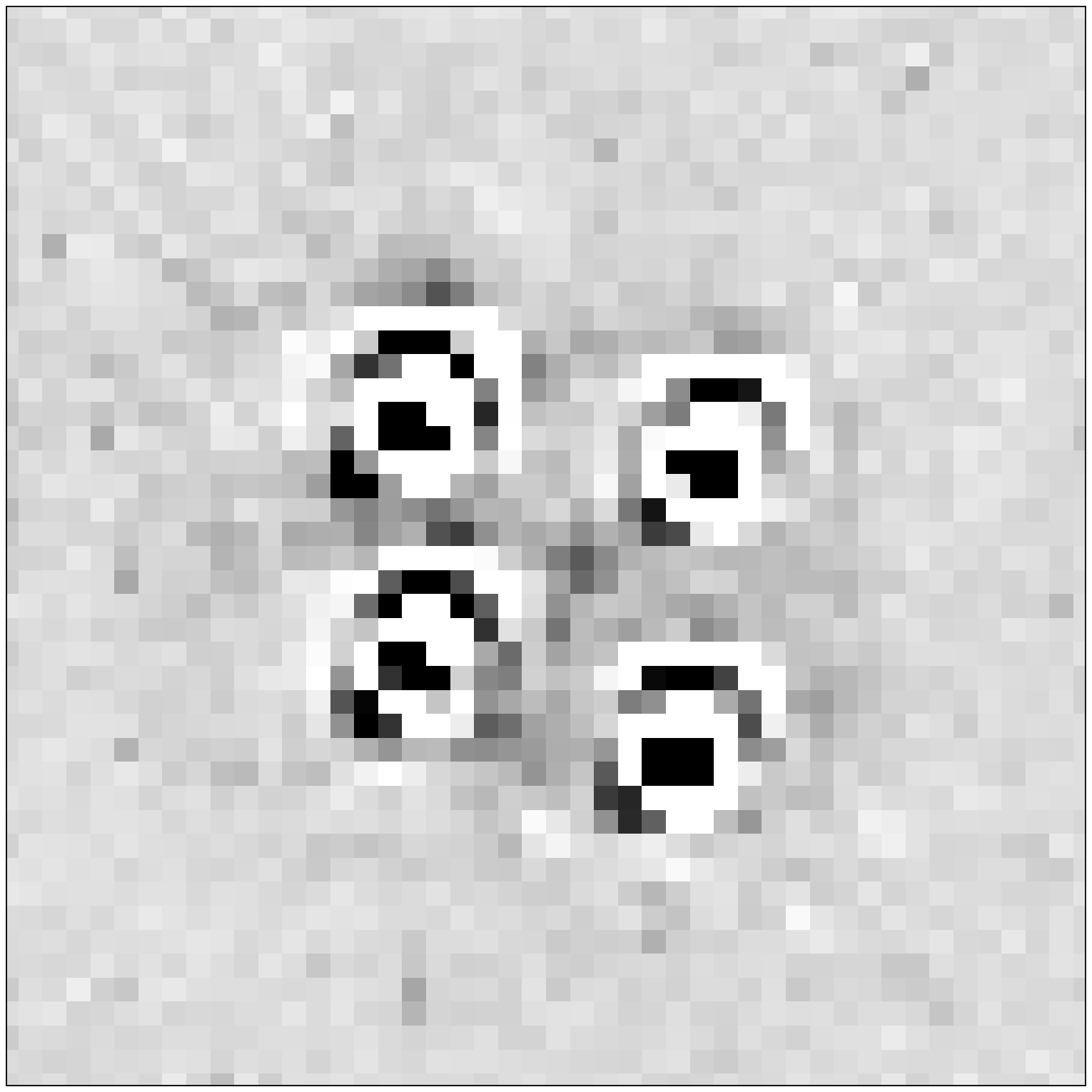,width=\textwidth}
\end{minipage}

\begin{minipage}{0.24\textwidth}
\psfig{file=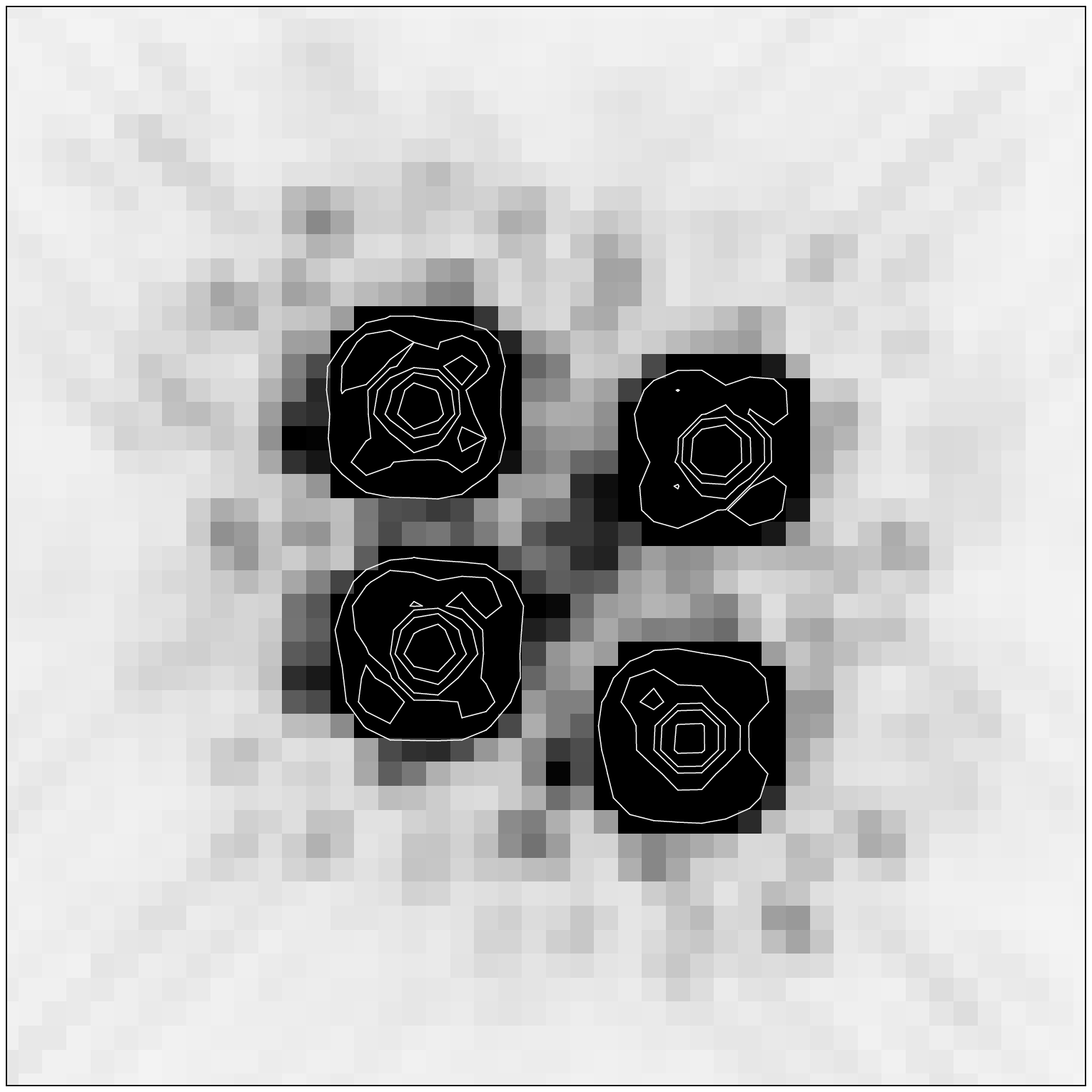,width=\textwidth}
\end{minipage}
\begin{minipage}{0.24\textwidth}
\psfig{file=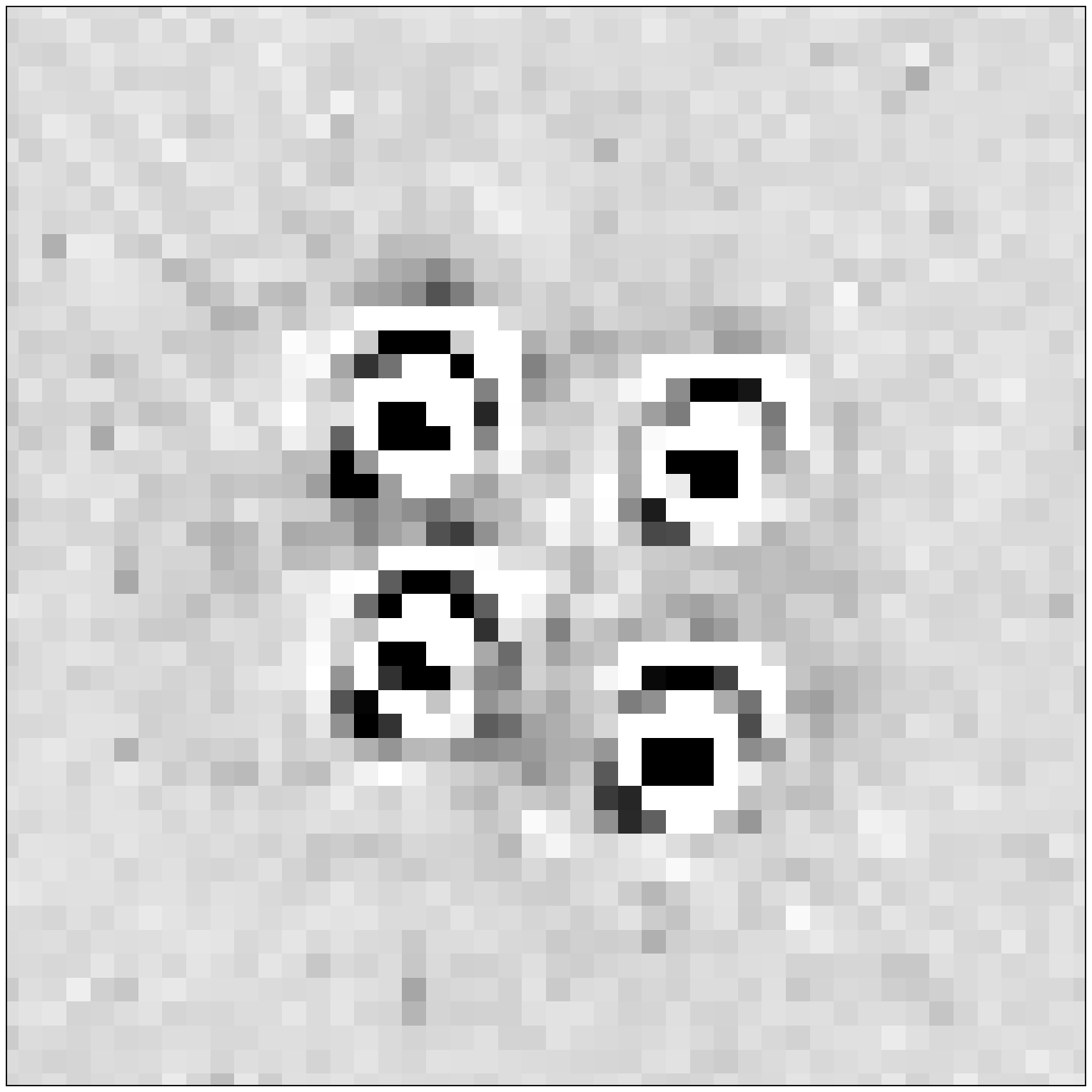,width=\textwidth}
\end{minipage}
\caption{
PSF subtraction steps, from left-to-right and top-to-bottom:
the NICMOS-2 observed image,
schematic of quasar-images and galaxy positions,
PSF model for the 4 quasars,
PSF subtraction of the 4 quasars,
PSF model for the 4 quasars plus the galaxy-lens,
PSF subtraction of the 4 quasars and the galaxy-lens. Size of these
rasters is 3.4 arcseconds on a side. The orientation corresponds to the
NICMOS-2 images (PA of image from Y axis 74.3765 degree).
\label{fig:psfsub}
}
\end{figure}

In order to quantify the magnitude and position of the galaxy-lens, 
we assume that its profile follows a de Vaucouleurs law and we include 
it in the fitting of the PSF subtraction. Parameters were chosen
in order to be left with a mean sky value 
similar to the median sky value. However we could not
constrain the structural parameters of the galaxy-lens due to poor
signal to noise.
Figure \ref{fig:psfsub} exhibits the different steps in the subtraction 
procedure.  Note that the galaxy-lens, $H1$, is clearly detected {\it even} 
on the non-PSF subtracted 
image. Its position and H$_{F160W}$ magnitude are given in Table \ref{tab:gal}.
Unfortunately, we can only get lower limits to its R$_{F702W}$ and I$_{F814W}$
magnitudes, preventing us from deriving a redshift constraint 
for H1. 

\begin{table*}
\begin{tabular}{lllllll}
\hline
ID & $\Delta\alpha$ & $\Delta\delta$ & 
U$_{F336W}$ & R$_{F702W}$ & I$_{814W}$ & H$_{160W}$ \\
 & (") & (") & 94/12/23 & 94/12/23 & 94/12/23 & 97/12/28  \\
\hline
A & 0.000  & 0.000 & 18.95        & 17.68      & 17.45	     & 15.67  \\
B & 0.744  & 0.172 & 19.31 [0.73] &17.82[0.88] & 17.58[0.89] & 15.76[0.92] \\
C & -0.491 & 0.716 & 19.93 [1.01] &17.95[0.77] & 17.73[0.76] & 14.03[0.71] \\
D & 0.355  & 1.043 & 19.13 [0.85] &17.97[0.74] & 17.81[0.70] & 14.29[0.57] \\
\hline
\end{tabular}
\caption{Relative positions (from Turnshek et al 1997),
photometry (U$_{F336W}$, R$_{F702W}$,I$_{F814W}$ \& 
H$_{F160W}$) and flux ratios of the 4 quasar images relative to image A.
Note the strong variation of the flux-ratio from U to H. 
\label{tab:qso}
}
\end{table*}

\section{The environment of the Cloverleaf:  photometric redshifts of
the faint objects in the field}

\begin{table*}

\begin{tabular}{lllllllll}
\hline
ID & $\Delta\alpha$ & $\Delta\delta$ & I$_{814W}$ & R$_{F702W}$ $-$ I$_{814W}$ 
& H$_{160W}$ & $z_{phot}$ & $\Delta z$ & Comments \\
 & (") & (") & & &  & max & $75 \%$ & \\
\hline
H1 & 0.12 & 0.50 & $>$22.5 & R $>$22.7 & 20.5$\pm$0.7 & -- & -- &
the position accuracy is $\pm$0.04 arcsec\\
H2 & -7.6 & -7.2 & $>$24.8 & R $>$25.5 & 21.6$\pm$0.7 & -- & -- &
very red, only detected in H.\\
\hline
1 & -7.3 & -2.1 & 22.52 & 0.72	& 20.7$\pm$0.3 & 0.8 & 0.6-1.1 &
E/S0 morphology\\
2 & -5.5 & -3.7 & 23.11 & 0.86	& 22.3$\pm$0.9 & 0.9 & 0.8-1.0 & \\
3 & -0.6 & 8.2 & 23.16 & 0.88	& 22.0$\pm$0.5 & 0.9 & 0.8-1.0 & 
disturbed morphology\\
7 & 3.9 & -5.5 & 23.75 & 0.74	& $>$22.3      & 0.9 & 0.5-1.2 & \\
8 & -0.7 & -4.1 & 23.78 & 1.16	& 21.8$\pm$0.5	& 1.0 & -- & $z_{phot}$
solution at $\sim 50 \%$ confidence level \\
9 & -5.2 & 2.2 & 23.94 & 0.12	& $>$22.3	& -- & -- & no  $z_{phot}$
constraint\\
11 & 3.2 & -6.8 & 24.01 & 1.09	& 22.3$\pm$0.8	& 1.0 & 0.7-1.2 & \\
14 & 4.7 & 0.2 & 24.37 & $>$1.1	& 20.0$\pm$0.2	& 2.0 & 1.8-2.7 & very red,
possibly associated with absorbers\\
15 & -2.2 & -7.8 & 24.41 & 1.05	& $>$22.3	& 0.8 & 0.6-1.2 & \\
19 & -0.9 & 5.3 & 24.74 & 0.88	& $>$22.3	& 1.0 & 0.5-1.5 & \\
20 &-6.1 &-6.1 & 24.78 & $>$0.7	& $>$22.3	& 1.1 & 0.2-1.6 & $z_{phot}$
is poorly determined \\
\hline
\end{tabular}

\caption{Relative position (to quasar image A), 
photometry (R$_{F702W}$,I$_{F814W}$ 
\& H$_{F160W}$), and predicted redshift of 
the galaxies within 10 arcsec around the Cloverleaf. The best matching
redshift is given, as well as the permitted redshift interval 
corresponding to a $75 \%$ confidence level.
\label{tab:gal}
}
\end{table*}

In the field-of-view covered by the H$_{F160W}$ image, we 
identify 12 faint objects for which we can 
provide at least one of the R$_{F702W}$, I$_{F814W}$ or
H$_{F160W}$ magnitude [in addition to H1]. Among these objects,
11 have been identified by Kneib et al (1998). Another one, called H2,
was not detected either on the R$_{F702W}$ or I$_{F814W}$ images. All these 
objects are shown on Figure \ref{fig:image}. 
Their positions with respect to the quasar image A are
given in Table \ref{tab:gal}, as well as their 
R$_{F702W}$, I$_{F814W}$ and H$_{F160W}$ magnitudes.

We have estimated the redshift of these objects from their
R$_{F702W}$, I$_{F814W}$ and H$_{F160W}$ magnitudes, according to
the standard minimization method described by Miralles \& Pell\'o (1998). 
The updated Bruzual \& Charlot evolutionary code (Bruzual \& Charlot, 
1993) was used and 5 different synthetic star-formation rates (SFR) tested: 
a 0.1 Gyr burst, a constant SFR, and 3 $\mu$ models 
(e-decaying SFR) with characteristic decay-time matching the sequence 
of colors for E, Sa and Sc
galaxies. For each SFR type, 51 template spectra were selected with
different stellar population ages. We also tested the stability 
of our results against different $A_V$ values in the galaxies. 
The template database includes 255 solar metallicity spectra. 

We have analyzed through simulations the accuracy and the 
possible biases affecting the redshift estimate from the photometry.
For this purpose, a simulated catalogue  of galaxies was created, 
with redshifts 
uniformly distributed between 0 and 6, and randomly sampling the SFR and ages. 
Photometric errors ( uncorrelated for the different filters)
were introduced as gaussian noise distributions with FWHM 
chosen so as to match the typical values for the Cloverleaf galaxy sample.
According to these simulations, the use of 
the R$_{F702W}$, I$_{F814W}$ and H$_{F160W}$ filters does not introduce any
systematic bias in the redshift determination for $0.5<z<2.5$. 

The most probable redshift was obtained for each object, as well as
its probability (in the sense of $\chi^2$) as a function of 
redshift (Table \ref{tab:gal}).
We find that 9 objects have a redshift between 0.8 and 1.1, with
$\overline{z} = 0.9 \pm 0.1$. Most of these objects are well fitted by 
a 1 to 2 Gyr burst models and $A_V=0$. The I$_{F814W}$ filter 
maps the B restframe at $z \sim 0.9$. The absolute magnitudes of these 
galaxies are -21$<M_B<$-19.

For 4 objects (including the galaxy-lens H1 and the very red object H2)
the redshift is poorly determined. The objects \#14 and H2 are extremely red in 
(I$_{F814W}$-H$_{F160W}$) and are consistent with galaxies at redshift 2 or 
higher. For object \#14, the best-fit model (without reddening),
is a 2.5 Gyr burst, which, given the redshift uncertainty,  leads to 
$M_B= -22.5$ to $-23.9$. This high luminosity may suggest
that object \#14 might as well be gravitationally amplified by the cluster-lens 
with $z\sim 1$.

The photometric redshifts were computed using a solar metallicity. 
This is a fair approximation as, for 
1$<z<$2 and the filters considered in this analysis, the results are rather 
unsensitive to metallicity. Furthermore, the estimated redshifts given in
Table \ref{tab:gal} are insensitive to a reddening $A_V < 2$. If the 
reddening is much higher,
the estimated redshift would shift toward a lower value. 
However there is no reason to believe that all objects in the Cloverleaf field 
should be highly reddened. 
The dispersion is high, however, as reflected in the large 
interval at the 75\% confidence level (Table 2); it is therefore 
important to effectively measure the redshift of the brightest
objects to validate this conclusion.

\section{Discussion and concluding remarks}

Two main results have been obtained from the NICMOS-2 data. For 
the first time the galaxy-lens, H1,
close to the line of sight toward the quasar, has been identified. Its position 
with respect to the quasar line-of-sight is found to be similar (within
the uncertainties) to the one derived in the various Cloverleaf 
gravitational lens models ({\it e.g.} Kneib et al, 1998).
It remains difficult to estimate the redshift of the galaxy-lens H1 because 
the PSF subtraction leaves an increased background noise in the region 
amid the 4 quasar images. Yet, a redshift estimate around 1.0 or higher
is consistent with the H$_{F160W}$
magnitude and the I$_{F814W}$ lower limit magnitude we have 
derived for H1. 
Clearly deep spectroscopic data are needed
to solve for the determination of its redshift.
We find also that there is a unique 
galaxy-lens, in contradiction to some early models in which two galaxy-lenses 
had been envisaged (Kayser et al, 1990).
Assuming that H1 is around $z\sim 1$, and has similar colors and
absolute magnitude than the nearby galaxies gives a 
Mass-to-light ratio of M($<5.1$kpc)/L$_B$ $\sim$ 25 M/L$_{B\odot}$.

With regard to the Cloverleaf environment, we show that 8 nearby galaxies 
have a most probable redshift around 0.9,
giving credit to the presence of a galaxy cluster/group along the line of
sight to the Cloverleaf.
In our previous modelling (Kneib et al, 1998),
we assumed for this galaxy cluster/group a redshift of 1.7, as a mean of the 
redshifts of the 4 absorbers silhouetted on the 
quasar spectrum. This value should be revised.
The location of the galaxy-lens 
is now known from the NICMOS-2 observations and will be implemented in a 
new model of the lens-system. One of the faint 
galaxies surrounding the Cloverleaf appears to be at a larger redshift, around 
2, and might be related with the absorber at z$=$2.07 or 2.09
(Monier et al 1998). 
Further IR imaging/spectroscopy of these galaxies should remove 
the remaining uncertainties of the Cloverleaf lens-system.
 
\acknowledgements{
We would like to thank G. Bruzual for allowing the use of his 
code as well as for useful discussions on photometric redshifts.
Many thanks, to Jens Hjorth for a careful reading of this manuscript
and fruitful discussion on lensing and other topics.
DA wishes to thank Observatoire Midi-Pyr\'en\'ees for hospitality.
}

\end{document}